\documentclass[prb,twocolumn,floatfix,superscriptaddress]{revtex4}
\usepackage{amssymb,amsmath,graphicx,afterpage}

\usepackage{natbib}
\usepackage[sort&compress]{natbib}
\makeatletter
\renewcommand*{\fnum@figure}{{\normalfont\bfseries \figurename~\thefigure}}
\makeatother
\usepackage{color,bm}
\pdfoutput=1

\begin{document}
\title{\boldmath Macroscopic Manifestation of Domain-wall Magnetism and Magnetoelectric Effect in a N\'eel-type Skyrmion Host \unboldmath}
\author{K. Geirhos$^*$}
\affiliation{Experimental Physics V, Center for Electronic
Correlations and Magnetism, University of Augsburg, 86135 Augsburg,
Germany}
\author{B. Gross}
\affiliation{Department of Physics, University of Basel, 4056 Basel,
Switzerland}
\author{B. G. Szigeti}
\affiliation{Experimental Physics V, Center for Electronic
Correlations and Magnetism, University of Augsburg, 86135 Augsburg,
Germany}
\author{A. Mehlin}
\affiliation{Department of Physics, University of Basel, 4056 Basel,
Switzerland}
\author{S. Philipp}
\affiliation{Department of Physics, University of Basel, 4056 Basel,
Switzerland}
\author{J. S. White}
\affiliation{Laboratory for Neutron Scattering and Imaging, Paul
Scherrer Institut, CH-5232 Villigen, Switzerland}
\author{R. Cubitt}
\affiliation{Institut Laue-Langevin, 6 rue Jules Horowitz, 38042
Grenoble, France}
\author{S. Widmann}
\affiliation{Experimental Physics V, Center for Electronic
Correlations and Magnetism, University of Augsburg, 86135 Augsburg,
Germany}
\author{S. Ghara}
\affiliation{Experimental Physics V, Center for Electronic
Correlations and Magnetism, University of Augsburg, 86135 Augsburg,
Germany}
\author{P. Lunkenheimer}
\affiliation{Experimental Physics V, Center for Electronic
Correlations and Magnetism, University of Augsburg, 86135 Augsburg,
Germany}
\author{V. Tsurkan}
\affiliation{Experimental Physics V, Center for Electronic
Correlations and Magnetism, University of Augsburg, 86135 Augsburg,
Germany} \affiliation{Institute of Applied Physics, Academy of
Sciences of Moldova, MD 2028, Chisinau, Republica Moldova}
\author{A. O. Leonov}
\affiliation{Center for Chiral Science, Hiroshima University, Higashi-Hiroshima, Hiroshima, 739-8526, Japan}
\affiliation{Department of Chemistry, Faculty of Science, Hiroshima University Kagamiyama, Higashi-Hiroshima, Hiroshima, 739-8526, Japan}
\author{S. Bord\'acs}
\affiliation{Department of Physics, Budapest University of
Technology and Economics, 1111 Budapest, Hungary}
\affiliation{Hungarian Academy of Sciences, Premium Postdoctor
Program, 1051 Budapest, Hungary}
\author{M. Poggio}
\affiliation{Department of Physics, University of Basel, 4056 Basel,
Switzerland}
\author{I. K\'ezsm\'arki}
\affiliation{Experimental Physics V, Center for Electronic
Correlations and Magnetism, University of Augsburg, 86135 Augsburg,
Germany}
\begin{abstract}
We report a magnetic state in GaV$_4$Se$_8$ which emerges
exclusively in samples with mesoscale polar domains and not in polar
mono-domain crystals. Its onset is accompanied with a sharp anomaly
in the magnetic susceptibility and the magnetic torque, distinct
from other anomalies observed also in polar mono-domain samples upon
transitions between the cycloidal, the N\'eel-type skyrmion lattice
and the ferromagnetic states. We ascribe this additional transition
to the formation of magnetic textures localized at structural domain
walls, where the magnetic interactions change stepwise and spin
textures with different spiral planes, hosted by neighbouring
domains, need to be matched. A clear anomaly in the magneto-current
indicates that the domain-wall-confined magnetic states also have
strong contributions to the magnetoelectric response. We expect
polar domain walls to commonly host such confined magnetic edge
states, especially in materials with long wavelength magnetic order.
\end{abstract}

\maketitle

Geometrical or dimensional constraints can promote the formation of
new quantum phases which are absent in unconstrained bulk systems.
Prominent examples include metallic surface states in topological
insulators~\cite{hasan2010colloquium}, superconducting vortex state
below the Kosterlitz-Thouless transition~\cite{resnick1981kosterlitz}, interface-induced 2D electron gas~\cite{ohtomo2004high} and superconductivity~\cite{qing2012interface,he2013phase,reyren2007superconducting}, integer and fractional quantum Hall edge
states~\cite{nakamura2019aharonov,cohen2019synthesizing,macdonald1990edge}
and Wigner crystals~\cite{andrei1988observation,shapir2019imaging}
in systems with reduced dimensions. Geometrical constraints are
usually enforced externally by synthesizing/structuring materials
with mesoscopic dimensions, e.g., in the form of thin films,
heterostructures, nanowires, quantum dots, etc. Such constraints can
also emerge naturally via the formation of mesoscale structural
domain patterns and topological
defects~\cite{zurek1985cosmological,choi2010insulating,chae2010self,
lilienblum2015ferroelectricity,ma2018controllable,seidel2016topological,
butykai2017characteristics,farokhipoor2014artificial}.

Indeed, structural domain walls (DW) have recently been reported to
possess novel functionalities, such as DW conductivity, electrical
rectification and super switching, as observed in
YMnO$_3$~\cite{choi2010insulating,ruff2017conductivity}, ErMnO$_3$~\cite{schaab2018electrical},
BiFeO$_3$~\cite{ma2018controllable,seidel2009conduction},
LiNbO$_3$~\cite{werner2017large},
Pb$_x$Sr$_{1-x}$TiO$_3$~\cite{matzen2014super}, etc. Besides the
peculiar electrical properties of DWs, the atomically sharp
structural changes associated with them can substantially modify the
magnetic exchange interactions and spin orders, as reported for
SrRuO$_3$--Ca$_{0.5}$Sr$_{0.5}$TiO$_3$ heterostructures and thin
films of La$_{2/3}$Sr$_{1/3}$MnO$_3$ and
TbMnO$_3$~\cite{kan2016tuning,liao2016controlled,farokhipoor2014artificial}.
Furthermore, geometrical constraints were shown to substantially
increase the thermal stability range of magnetic
skyrmions~\cite{yu2011near,sonntag2014thermal,mehlin_stabilized_2015}, which are whirling
spin textures on the nanoscale, and to generate exotic magnetic edge
states, such as chiral bobbers~\cite{zheng2018experimental}.

Lacunar spinels with the chemical formula AB$_4$X$_8$ (A = Al, Ga,
Ge; B= V, Mo, Nb, Ta; X = S, Se, Te) are a family of narrow-gap Mott
insulators, exhibiting a plethora of correlation and spin-orbit
effects, including pressure-induced
superconductivity~\cite{abd2004transition}, bandwidth-controlled
metal-to-insulator transition~\cite{phuoc2013optical},
electric-field-driven avalanche of the Mott
gap~\cite{guiot2013avalanche}, large negative
magnetoresistance~\cite{dorolti2010half}, two-dimensional
topological insulator state~\cite{kim2014spin} and orbitally driven
ferroelectricity~\cite{singh2014orbital,wang2015polar,geirhos2018orbital,ruff2017polar}.
In addition to these charge related phenomena and particularly
important for the present study, lacunar spinels, such as
GaV$_4$S$_8$, GaV$_4$Se$_8$ and GaMo$_4$S$_8$, were the first
material family found to host the N\'eel-type skyrmion lattice (SkL)
state~\cite{kezsmarki2015neel,bordacs2017equilibrium,butykai2019squeezingmagnetic}.
In the last two compounds, the SkL state was reported to be stable
down to zero Kelvin. While these materials have a cubic structure
($F\overline{4}3m$) at high temperature, many of them transform to a
rhombohedral state ($R3m$) upon a Jahn-Teller transition, occuring
around 30-50\,K, which is triggered by the degeneracy of the
B$_4$X$_4$ cluster
orbitals~\cite{pocha2000electronic,wang2015polar,hlinka2016lattice}.
The polar axial symmetry of the low-temperature state is a
prerequisite for the emergence of the N\'eel-type SkL
state~\cite{bogdanov1989thermodynamically}.

In this work, we demonstrate that non-trivial spin textures exist not only in the bulk, but complex magnetic states arise on the polar DWs of GaV$_4$Se$_8$: The
matching of spin cycloids at polar DWs gives rise to the emergence
of a novel magnetic state confined to the DWs.

\section*{Results}
\textbf{Electric control of polar domains.}

In these compounds, four possible polar domains with polarizations along the cubic $\langle$111$\rangle$-type axes can exist, that naturally form multi-domain, lamellar patterns. The DWs are \{100\}-type planes with typical periodicity in the range of 10-1000\,nm, as
observed in GaV$_4$S$_8$~\cite{butykai2017characteristics},
GaV$_4$Se$_8$~\cite{milde2019unpublished} and
GaMo$_4$S$_8$~\cite{neuber2018architecture} using atomic and
piezoelectric force microscopy.

\begin{figure}
\includegraphics[width=\linewidth] {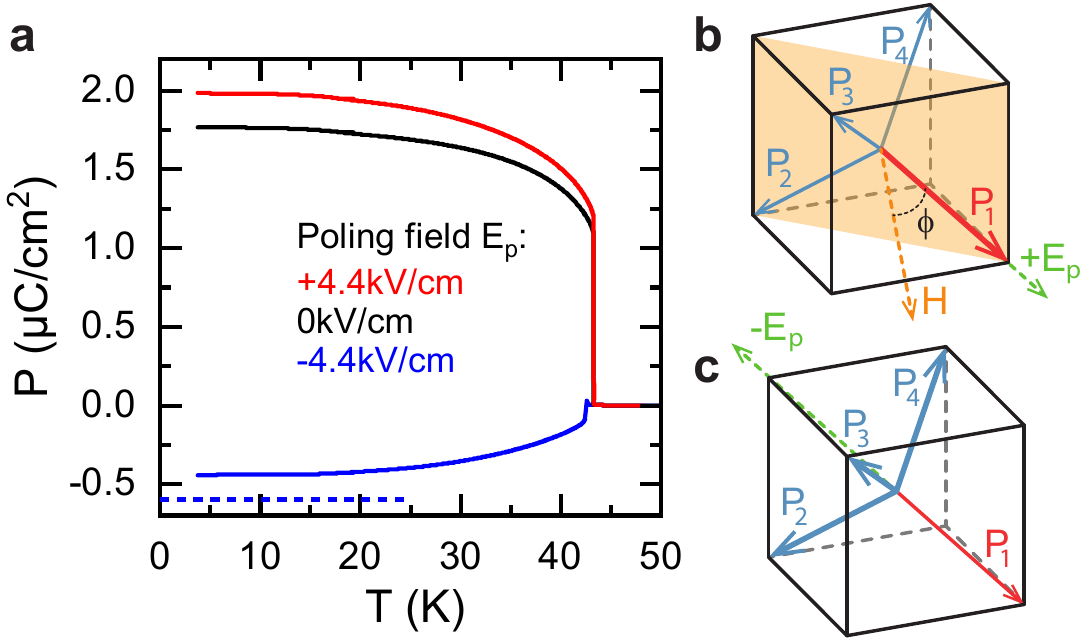}
\caption{\textbf{Electric control of polar domains in
GaV$_4$Se$_8$.} \textbf{a,} Temperature dependence of the
polarization along the [111] axis, measured during heating after
poling with different electric fields, $E_p$ $\parallel$[111], upon
cooling. The dashed line indicates the saturation polarization
expected for negative poling fields. \textbf{b \& c,} Schematic
representations of the cubic unit cell indicating the four possible
directions of the polarization, $P_{1-4}$, emerging in the
rhombohedral phase. For +$E_p$, the unique $P_1$ domain is favored
(\textbf{b}), for -$E_p$, the $P_{2-4}$ domains are favored
(\textbf{c}). In the magneto-current and torque experiments,
described later, the magnetic field was rotated approxiamtely in the
(1$\overline{1}$0) plane, highlighted in orange in panel \textbf{b}.
The angle $\phi$ is spanned by the field and the [111] axis.}
\end{figure}

Besides the high DW density, another peculiarity of this polar phase
is the lack of 180$^{\circ}$ DWs and the presence of 109$^{\circ}$
DWs only, owing to the non-centrosymmetric nature of the cubic
phase~\cite{butykai2017characteristics,neuber2018architecture}.
(This is in contrast to perovskite ferroelectrics, such as
BiFeO$_3$, where the paraelectric cubic phase is centrosymmetric.) Due to the lack of inversion domains, electric fields of opposite
signs, applied along one of the four $\langle$111$\rangle$-type
axes, select either the unique domain with polarization parallel to
the field or the other three domains, whose polarizations span
71$^{\circ}$ with the electric field. Such control of the polar
domains in GaV$_4$Se$_8$ is sketched in Fig.~1 and also demonstrated
via the temperature-dependent polarization, which was determined
from pyrocurrent measurements, following electric-field poling.
Indeed, the maximum polarization reached via positive electric
fields (2.0\,$\mu$C/cm$^2$) is considerably larger than for negative
fields (-0.45\,$\mu$C/cm$^2$). The ratio of the polarization values
in the polar mono- and multi-domain states is expected to be 3:1 in
the previous scenario, since the pyrocurrent component along the $P_1$ polar axis was detected.

\begin{figure}
\includegraphics[width=\linewidth] {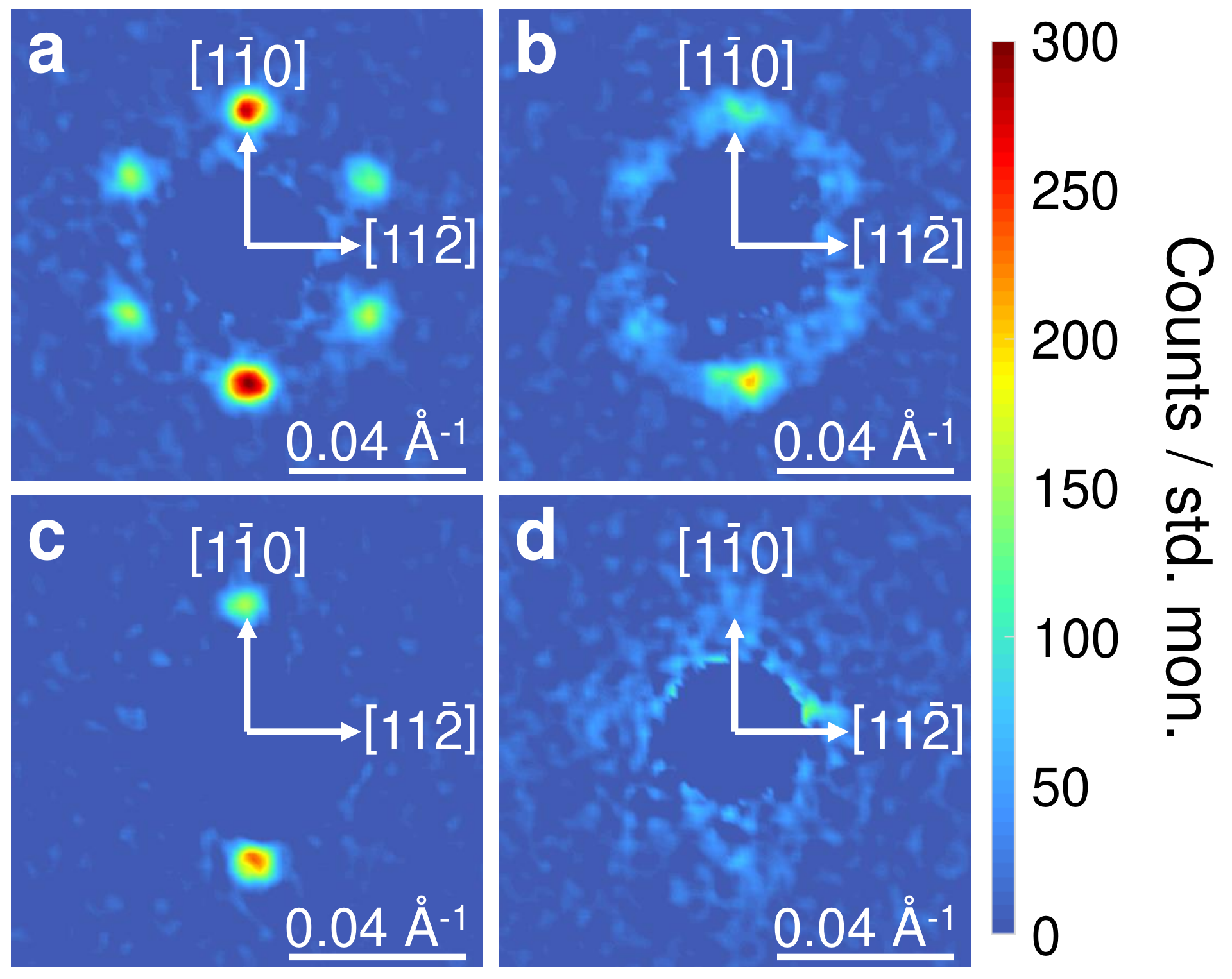}
\caption{\textbf{Variation of the magnetic structure of
GaV$_4$Se$_8$ in oblique magnetic fields.} \textbf{a-d,} SANS images
recorded at 12\,K in 220\,mT with the magnetic field spanning
0$^\circ$, 21$^\circ$, 27$^\circ$ and 35$^\circ$ with the [111]
axis, respectively. The hexagonal and the $\pm q$ patterns in panels
\textbf{a} and \textbf{c} correspond to the SkL and cycloidal
(conical) states, respectively. Panels \textbf{b} and \textbf{d} are
images taken near the SkL$\rightarrow$cycloidal and
cycloidal$\rightarrow$FM transitions, respectively.}
\end{figure}

\textbf{Magnetic states in polar mono- and multi-domain
GaV$_4$Se$_8$ crystals.}

The different polar domains are also distinguished magnetically,
since they are characterized by different directions of the uniaxial
anisotropy, coinciding with the polar
axes~\cite{ehlers2017exchange,ehlers2016skyrmion,bordacs2017equilibrium},
and by different orientations of the Dzyaloshinskii-Moriya vectors that prefer modulated magnetic structures with q-vectors perpendicular to the polar axis.
Correspondingly, for arbitrary directions of the magnetic field,
distinct magnetic states can coexist on the different types of polar
domains, as has been imaged in GaV$_4$S$_8$ by magnetic force
microscopy~\cite{kezsmarki2015neel}. Furthermore, twisted magnetic
textures confined to the vicinity of polar DWs have been observed in
the same material~\cite{butykai2017characteristics}, though a
systematic study of these edge states with changing the orientation
of the field could not be performed.

Due to the uniaxial anisotropy of the material, the magnetic state
depends not only on the magnitude, but also on the orientation of the
field. A sequence of three different phases,
SkL$\rightarrow$cycloidal$\rightarrow$ferromagnetic (FM), was observed at 12\,K by
small angle neutron scattering (SANS) while rotating a magnetic
field of $\mu_0 H=220$\,mT in the (1$\overline{1}$0) plane from the polar [111] axis
($H_{\parallel}$) towards the perpendicular plane ($H_{\perp}$), as
demonstrated in Fig.~2. The SANS patterns were recorded on the (111)
plane for several angles of the field.

\begin{figure*}
\includegraphics[width=\linewidth] {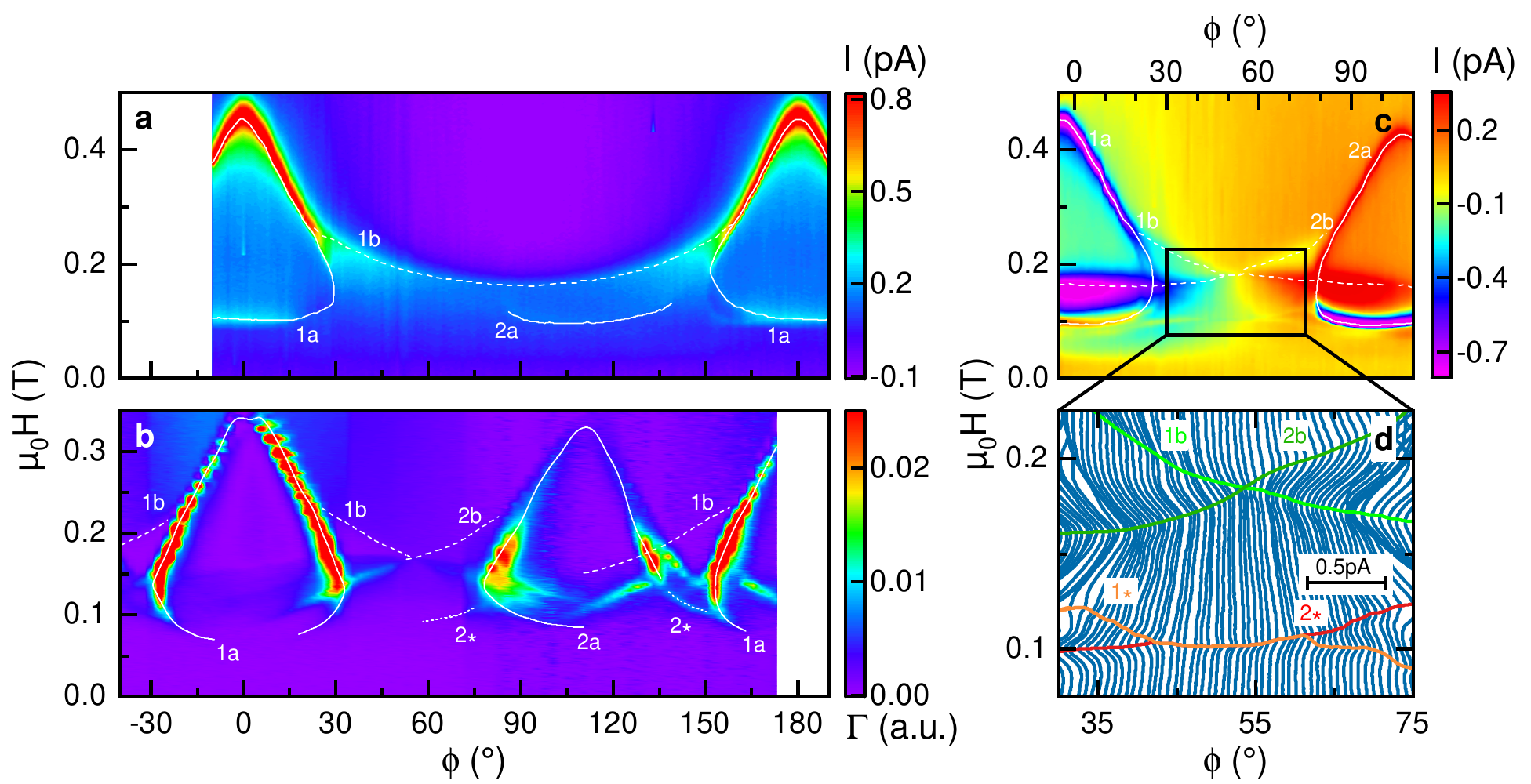}
\caption{\textbf{Angular dependence of the magneto-current
and the magnetic torque in GaV$_4$Se$_8$ at 12\,K.} Colour maps of
\textbf{a \& c,} the magneto-current and \textbf{b,} the dissipative
part of the torque, as functions of the angle ($\phi$) and the
magnitude ($H$) of the magnetic field. Data shown in panel \textbf{a} were
recorded on a nearly mono-domain sample, while data shown in panels \textbf{b \& c} were obtained on polar multi-domain samples. Panel \textbf{d} shows a magnified view of the frame
indicated in panel \textbf{c}, as a waterfall diagram, where the
magneto-current curves recorded at different angles are shifted
horizontally in proportion to $\phi$. In each panel, the anomalies
which can clearly be assigned to magnetic phase boundaries are
highlighted by white lines and labeled, as follows. The numbers
\textit{1} and \textit{2} correspond to domains with polar axes
[111] and [$\overline{1}\overline{1}$1], respectively. The characters
\textit{a} and \textit{b} correspond to the phase boundary of the
SkL state (SkL--cycloidal and SkL--FM) and the cycloidal--FM
boundary, respectively. The asterisks in panels \textbf{b} \&
\textbf{d} mark the newly found anomaly, associated with a magnetic
transition within the DWs.}
\end{figure*}

\begin{figure}[h!]
\includegraphics[width=\linewidth] {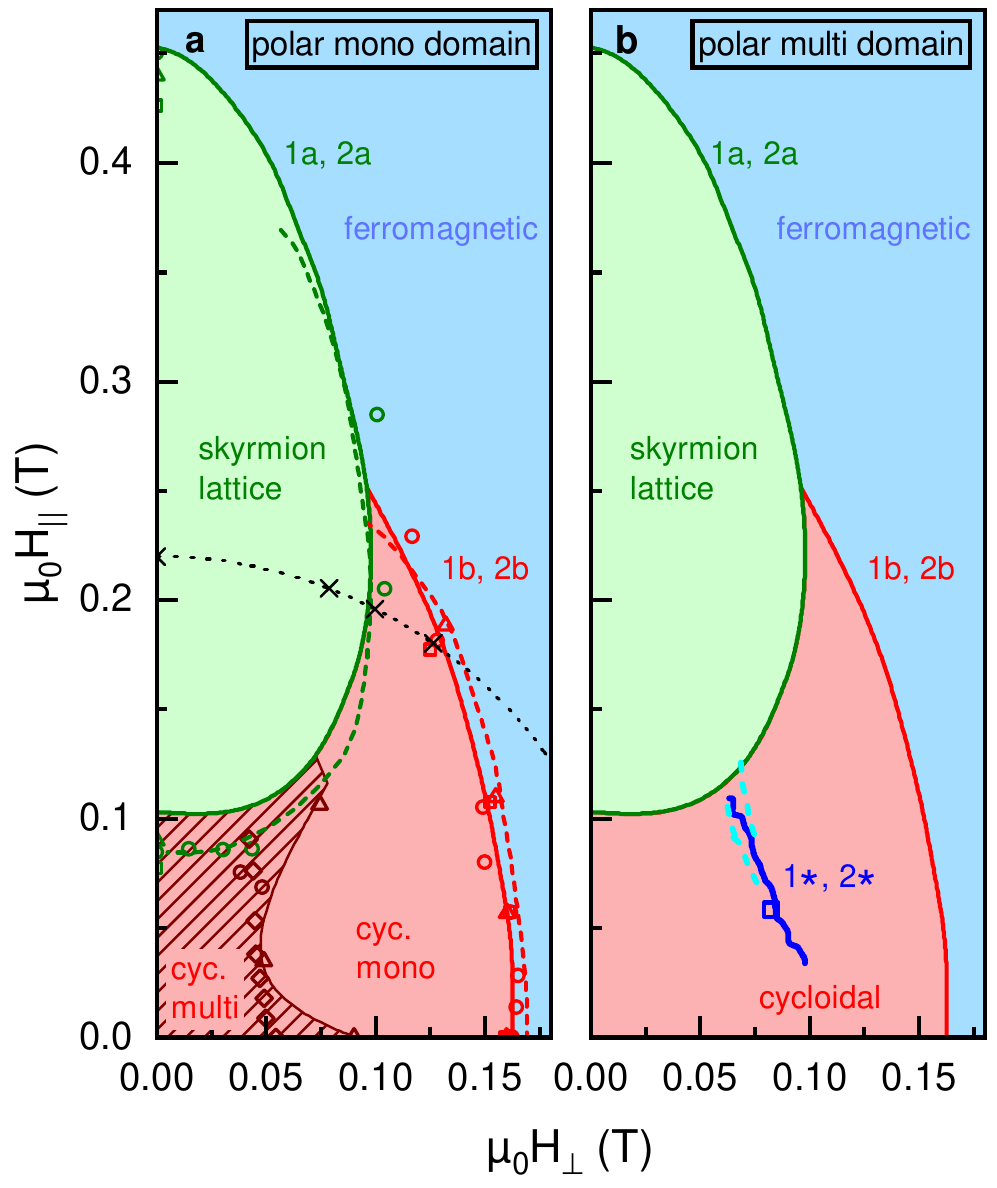}
\caption{\textbf{Magnetic phase diagram representative to
polar mono- and multi-domain GaV$_4$Se$_8$ crystals at 12\,K,
displayed in panel \textbf{a \& b}, respectively.} The solid and
dashed lines are phase boundaries deduced from magneto-current and
torque measurements, respectively. Triangles represent SANS data,
while circles/squares correspond to magnetization measurements
without/with electric field poling. Note that panel \textbf{a} does
not exclusively contain data obtained on polar mono-domain samples,
but includes all anomalies, which are present irrespective if the
samples are mono- or multi-domain ones. In case of polar multi-domain crystals, $H_\parallel$ and $H_\perp$ refers to field components parallel and perpendicular to the polar axis of the domain to which the anomalies are assigned. For the assignment of the
different anomalies to different polar domains in multi-domain
crystals and the establishment of the unified phase diagram, see the
main text. The hashed low-field area in panel \textbf{a} indicates
the region of the cycloidal q-vector reorientation, while the black
crosses along the dashed black line correspond to the points where
the SANS images in Fig.~2 were taken. In panel \textbf{b}, the phase
boundaries, which are also present in the polar mono-domain case,
are shown only as deduced from the magneto-current measurements.
Labels of the phase boundaries (\textit{1a}, \textit{1b},
\textit{2a}, \textit{2b}) correspond to those in Fig.~3. The additional phase boundary labeled by \textit{1$*$} \& \textit{2$*$} is determined by magneto-current (solid blue line), by torque (dashed light blue line) and by susceptibility measurements (blue square).}
\end{figure}

In order to continuously follow the phase boundaries on the
$H_{\parallel}$--$H_{\perp}$ plane, we carried out magneto-current
and magnetic torque measurements, when the magnetic field was
rotated in fine steps within the ($1\overline{1}$0) plane between
successive field sweeps. The corresponding data are depicted in
Fig.~3 in the form of colour maps over the field
magnitude--orientation plane ($\phi$ denotes the angle of the field
spanned with the [111] axis, as sketched in Fig.~1b), where the
magnitude of the magneto-current (Figs.~3a \& c) and the dissipative
part of the torque signal (Fig.~3b) are represented by colours. The
anomalies labeled in Fig.~3 are used to establish the magnetic phase
diagrams corresponding to polar mono- and multi-domain samples in
Figs.~4a \& b, respectively, also supplemented with magnetization,
susceptibility and SANS measurements. The phase boundaries
determined from the torque measurements slightly deviate from those
deduced by other methods, which is attributed to differences in the
measurements conditions (sample temperature, sample orientation, direction of the field
sweep, etc.). For a better match, the phase boundaries determined
from torque measurements were plotted in Fig.~4a with $H$ values
rescaled by a factor of $\sim$1.12.

Fig.~3a shows magneto-current studies on a nearly polar mono-domain
crystal, where all the observed anomalies (labeled as \textit{1a}
and \textit{1b}) are assigned to transitions in the $P_1$ domain
with polar axis along the [111] direction, except for a faint signal
(labeled as \textit{2a}) in the angular range of 85-145$^{\circ}$. The
180$^{\circ}$ periodicity of the signal can be readily followed. In
contrast, Fig.~3b shows magnetic torque data as obtained on a polar
multi-domain sample. Here, the faint feature, which is hardly
traceable in the magneto-current data on the mono-domain sample,
shows up clearly as a replica of the droplet-like motifs, which are
the strong features centered around 0$^{\circ}$ and 180$^{\circ}$ in
the nearly mono-domain sample. This additional droplet is also found
in magneto-current studies on multi-domain samples, as shown in
Fig.~3c. It is centered at around $\sim$109$^{\circ}$, the angle
spanned by the polar axes [111] and [$\overline{1}$ $\overline{1}$1].
Thus, it originates from the $P_2$ polar domain, whose polar axis
also lies in the (1$\overline{1}$0) rotation plane of the field. For
multi-domain samples, the detailed angular dependence of the
magneto-current and torque data reveal which magnetic anomalies can
be assigned to $P_1$ or $P_2$ polar domains, and help to establish a
domain-specific magnetic phase diagram (Fig.~4a), as if the samples
were polar mono-domain. Magnetic anomalies corresponding to $P_3$
and $P_4$ domains cannot be unambiguously assigned, mainly because
the angle spanned by the magnetic field and their polar axes varies
over a limited range, 54-90$^{\circ}$, while the field is rotated in
the (1$\overline{1}$0) plane.

In polar mono-domain samples (Fig.~4a), there are three magnetic
phases showing up, the cycloidal, the N\'eel-type SkL and the FM
state, in accord with former
studies~\cite{bordacs2017equilibrium,fujima2017thermodynamically}.
Note that phase boundaries in Fig.~4a are displayed from one hand, as obtained from magneto-current data on a nearly polar mono-domain crystal (Fig.~3a). On the other hand, Fig.~4a also shows how the phase diagram looks like for a given domain in a polar multi-domain crystal, when the orientation of the field is measured from the polar axis of that domain. Since there is no difference between the two cases, phase boundaries in Fig.~4a are representative to the magnetic states inside the polar domains. In fields applied along the polar axis, the cycloidal phase has
coexisting domains with wavevectors oriented in different directions
(in the plane perpendicular to the rhombohedral axis) up to its
transformation to the SkL
state~\cite{kezsmarki2015neel,bordacs2017equilibrium}. By oblique
fields, this cycloidal multi-domain state is turned to a cycloidal
mono-domain state, since $H_{\perp}$ selects the unique q-vector,
which is perpendicular to this field
component~\cite{bordacs2017equilibrium}. This repopulation of the
cycloidal domains is directly detected by SANS experiments (see
Fig.~S1 in the Supplemental material) and also manifested in
magnetic susceptibility and magneto-current data as a low-field
upturn and a peak, respectively, at $H_{\perp} \sim 40\,mT$, as
discerned e.g. in Figs.~5a \& b. (For details see Fig.~S2 in the
Supplemental material.) The crossover regime from the cycloidal
multi- to mono-domain state is also indicated in Fig.~4a. Another
important observation is that for certain angles of oblique fields,
there is a re-entrant cycloidal state, where the full sequence of
magnetic transitions with increasing field is
cycloidal$\rightarrow$SkL$\rightarrow$cycloidal$\rightarrow$FM. This
re-entrance of the cycloidal phase, found in magnetization,
magneto-current and torque experiments for $\phi=21-29^{\circ}$, has
been predicted for oblique fields in
GaV$_4$Se$_8$~\cite{leonov2017skyrmion}, but has not been observed
yet~\cite{bordacs2017equilibrium,fujima2017thermodynamically}.

In polar multi-domain samples, besides the features assigned to
magnetic transitions within the bulk of the different polar domains, there are
additional anomalies in the torque and magneto-current data, marked
by asterisks in Figs.~3b \& d, respectively. In case of
magneto-current measurements, the corresponding angular range
($\phi$ $=$30-75$^{\circ}$) is highlighted by a rectangle in Fig.~3c
and a zoom-in view of the magneto-current curves is given in
Fig.~3d. In the torque data, there are additional weak features observed~\cite{poggio2019unpublished}. However, here we only discuss anomalies simultaneously discerned in magnetization, magneto-current and torque data.

The key observation of the present work is the emergence of this
additional magnetic transition in polar multi-domain samples, as
shown in Fig.~4b. This change in the magnetic state is triggered by
oblique fields and located in the middle of the cycloidal phase,
where the cycloidal modulation with a unique q-vector is already
established by the $H_{\perp}$ component of the field. Fig.~5 summarizes the signatures of this
extra transition, based on various quantities measured in
$\mathbf{H}$ $\parallel$ $[001]$. This choice of the field orientation
guarantees that all the polar domains host the same magnetic state,
irrespective of the magnitude of the field, since the [001] axis
spans the same angle (54$^{\circ}$) with the magnetic anisotropy
axes of all the four polar domains. Thus, one would expect the same
magnetic anomalies, irrespective of having a polar mono- or
multi-domain sample, unless the presence of DWs introduces an
additional magnetic state not existing inside the domains. The magnetic susceptibility has a sharp peak at
$\sim$100\,mT for the polar multi-domain sample, which is strongly
suppressed as the polar mono-domain state is approached, but not fully achieved by poling. The poling worked more efficiently for the thin platelet used in the magneto-current studies: The sharp step in the magneto-current curve associated with this transition is completely suppressed, as the polar mono-domain state is reached. The dissipative part of the magnetic torque signal also shows
a weak anomaly at this field, though it is better visible a few
degrees away from the [001] direction. In this case the mono-domain
state could not be studied due to the unfeasibility of electric
poling. In the field dependence of the SANS intensity we could not
observe any anomaly associated with this transition. This may be due
to the signal to noise ratio and the limited field resolution in the
SANS measurements.

\begin{figure}
\includegraphics[width=\linewidth] {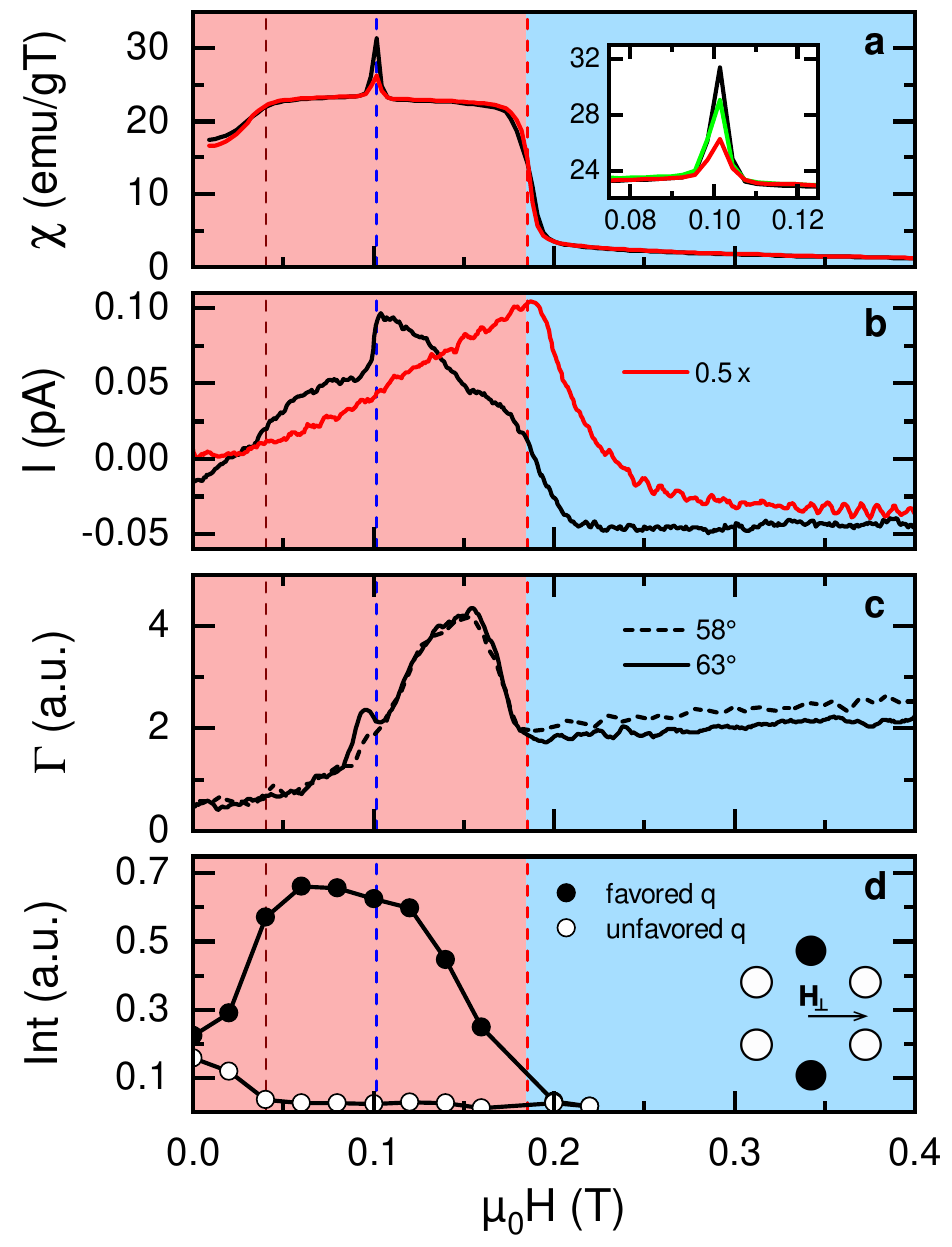}
\caption{\textbf{Magnetic anomalies in polar mono- and
multi-domain GaV$_4$Se$_8$ crystals at 12\,K.} Magnetic field
dependence of \textbf{a,} the real part of the ac-susceptibility,
\textbf{b,} the magneto-current, \textbf{c,} the dissipative part of
the torque and \textbf{d,} the SANS intensity, all measured in
$\mathbf{H}\parallel$ [001] ($\phi \sim 54^\circ$). Curves displayed
in black/red were recorded on polar multi-/mono-domain crystals. The
inset in panel \textbf{a} shows a magnified view of the anomaly
around 100\,mT for three different electric poling fields (red:
+2.5\,$\frac{\mathrm{kV}}{\mathrm{cm}}$, green:
0\,$\frac{\mathrm{kV}}{\mathrm{cm}}$, black:
-2.5$\frac{\mathrm{kV}}{\mathrm{cm}}$). The brown, blue and red
vertical dashed lines indicate the q-vector re-orientation, the
anomaly observed in multi-domain samples and the
cycloidal$\rightarrow$FM transition, respectively.}
\end{figure}

Another observation supporting that the new anomaly is associated
with magnetic states confined to the vicinity of DWs and not
extended over the entire volume of the crystals is the weakness of
the anomaly in the magnetic properties. Namely, the magnitude of the
step-wise increase in the magnetization, corresponding to the
susceptibility peak observed at 100\,mT with
$\mathbf{H}$$\parallel$[001], is less than 5\% of the magnetization
step observed upon the cycloidal to SkL transition in similar fields
for $\mathbf{H}$$\parallel$[111]. (See Fig.~S3 in the Supplemental
material.) Another possibility would be that while the anomaly is
readily linked to the presence of DWs, as demonstrated by electric
poling experiments, it is associated to a novel magnetic state
triggered by the DWs but not confined to their vicinity. However, in
this case the critical field of the corresponding transition would
depend on the distance of neighbouring DWs and, therefore, should
show a broad distribution. In contrast, the observed anomaly is
rather sharp with a full width of $\lesssim$3\,mT.

\section*{Discussion}

Our study demonstrates that the existence of DWs is a prerequisite
for the emergence of the new magnetic state in GaV$_4$Se$_8$, as
revealed by macroscopic magnetic and magnetoelectric properties. The
density of DWs is rather high in this
compound~\cite{milde2019unpublished}, especially if we compare the
typical DW distance of $\sim$100-200\,nm to the wavelength of the
N\'eel-type magnetic modulations of $\sim$20\,nm. The DWs separating
polar rhombohedral domains in this compound do not only produce a
sudden change in the direction of the magnetic anisotropy axis but
also change the orientation of the Dzyaloshinskii-Moriya vectors. We
think that the mismatch between different modulated states, favoured
by different magnetic interactions at adjacent domains, can give
rise to the formation of new spin textures at the DWs. We expect that the new magnetic state conﬁned to DWs transforms to the FM state at the same time as the cycloidal to FM transition happens inside the domains. This is in accord with the lack of a second weak anomaly in Fig.~4b, based on the magnetization, torque and magneto-current data for polar multi-domain samples. An alternative scenario is that the magnetic frustration coming from
the change of the Dzyaloshinskii-Moriya interaction at DWs may also
suppress the stability range of the modulated states near the DWs,
since the critical magnetic field, where the modulated states
transform to the FM state, scales with $D^2$/$J$, where $D$ is the strength
of the Dzyaloshinskii-Moriya interaction and $J$ is the isotropic
Heisenberg exchange. The corresponding scenario for
the magnetic anomaly, observed exclusively in polar multi-domain
samples, is that the DW region becomes FM at lower fields, when the
cycloidal modulations still persist inside the domains.

It is important to note that the two transition lines \textit{1$*$} and \textit{2$*$} in Figs.~3c \& d fall on each other and form a single phase boundary in Fig.~4b, if they are respectively assigned to $P_1$ and $P_2$ domains, meaning that the angle of the magnetic field is measured from the corresponding polar axes. From this, we conclude that these conﬁned magnetic states do not form on DWs between $P_1$ and $P_2$ domains, but on $P_1-P_3$ or $P_2-P_4$ DWs parallel to the (100) plane or on $P_1-P_4$ or $P_2-P_3$ DWs parallel to the (010) plane. As discussed earlier,
the magnetic anisotropy axis of each domain spans the same angle with
the field for $\mathbf{H}$ $\parallel$[001], thus, the same magnetic
state exists inside all the domains. In contrast, not all the DWs
are magnetically equivalent for this specific field
orientation: $\mathbf{H}$ $\parallel$[001] lies within the plane of
the four types of DWs listed above and is perpendicular to the
$P_1-P_2$ and $P_3-P_4$ DWs. (For details, see Fig.~S4 in the
Supplemental material.) Note that in the angular range, where the
new state appears at the DWs, the corresponding domains host the
cycloidal state. Consequently, the edge state emerges due to the
imperfect matching of two cycloids with different rotation planes.
A similar phenomenon has recently been observed for antiferromagnetic
spin cycloids interfaced at polar DWs of
BiFeO$_3$~\cite{chauleau2019electric,xue2019electricfieldinduced}.
More specific information about the nature of the new magnetic state
would require the real-space imaging of magnetic textures in
GaV$_4$Se$_8$. Such trials on GaV$_4$Se$_8$ have not been successful
so far, in contrast to GaV$_4$S$_8$, where we could even observe
magnetic textures localized to DWs~\cite{butykai2017characteristics}
besides the extended cycloidal and SkL
states~\cite{kezsmarki2015neel}.

Recently, the influence of structural DWs on macroscopic (bulk)
electronic properties has been extensively investigated~\cite{werner2017large,ruff2017conductivity}. In contrast, manifestations of spin textures confined to DWs in
macroscopic magnetic properties have hardly been observed. In the
present work, we found sharp magnetic and magnetoelectric anomalies
in GaV$_4$Se$_8$ what we assign to a phase transition originating
from magnetic states confined to the vicinity of structural DWs.

\section*{Methods}

\textbf{Sample synthesis and characterization.} Single crystals of
GaV$_4$Se$_8$ with typical mass of 1-30\,mg were grown by the
chemical vapour transport method using iodine as the transport
agent. The crystallographic orientation of the samples was
determined by X-ray Laue and neutron diffraction.

\textbf{Magnetization measurements.} The magnetization and ac
susceptibility measurements were performed following an initial
zero-field cooling, using an Magnetic Property Measurement System
(MPMS) from Quantum Design. For ac susceptibility measurements with electric field poling, a home made probe was used. Poling electric field was applied only during cooling.

\textbf{Magneto-current measurements.} For the pyro- and
magneto-current measurements, contacts on two parallel (111) faces of
the crystals were applied using silver paint. The sample was cooled
using an Oxford helium flow cryostat. For changing the orientation
of the magnetic field with respect to the electric contacts, the
sample was mounted on a platform, which can be rotated by a stepper
motor. To ensure low noise measurements, the platform was equipped
with two coaxial cables. For magneto-current measurements, the
current was recorded using Keysight B2987A electrometer while the
magnetic field was swept with a typical rate of 0.5-1\,T/min. For
pyrocurrent measurements, the current was recorded with the same
electrometer while the temperature was swept from 3\,K to 50\,K with
a rate of about 6\,K/min.

\textbf{Torque magnetometry.} Torque data was recorded with dynamic
cantilever magnetometry
(DCM)~\cite{mehlin_stabilized_2015,gross_dynamic_2016}. A single
crystalline piece of GaV$_4$Se$_8$ (spacial dimensions in the order
of 100\,$\mu$m) was attached to the end of a commercial cantilever
(Arrow-TL1, Nanoworld). In DCM, the cantilever is driven into
self-oscillation at its resonance frequency. Changes in the
dissipation $\Delta\Gamma = \Gamma-\Gamma_0$, induced by the torque
from the sample, are measured as a function of the uniform applied
magnetic field $\mathbf{H}$, where $\Gamma_0$ is the cantilever's
intrinsic mechanical dissipation at $H = 0$. Measurements of
$\Delta\Gamma$ are particularly useful for identifying magnetic
phase transitions~\cite{mehlin_stabilized_2015}.

DCM measurements were carried out in a vibration-isolated
closed-cycle cryostat. The pressure in the sample chamber is less
than $10^{-6}$\,mbar and the temperature can be stabilized between
4-300\,K. Using an external rotatable superconducting magnet,
magnetic fields up to 4.5\,T can be applied along any direction
spanning 120$^{\circ}$ in the plane of cantilever oscillation. A
piezoelectric actuator mechanically drives the cantilever with a
constant oscillation amplitude of a few tens of nanometers using a
feedback loop implemented by a field-programmable gate array. The
cantilever's motion is read out using an optical fiber interferometer
using 100\,nW of laser light at 1550\,nm~\cite{rugar_improved_1989}.

\textbf{Small-angle neutron scattering.} Small-angle neutron
scattering (SANS) was performed on a 11.6\,mg single crystal sample
of GaV$_4$Se$_8$. SANS patterns discussed in the main text were
measured using the SANS-I instrument of SINQ at the Paul Scherrer
Institut (PSI), Villigen, Switzerland. The wavelength of the
incoming neutrons was 6\,\AA, the collimation and the
sample-detector distances were set to 6\,m. The magnetic field was
applied in various directions within the (1$\overline{1}$0) plane.
Additionally, magnetic field dependence of the scattering data shown
in the supplement was measured using the D11 and the D33 instruments
at the Institut Laue-Langevin (ILL), Grenoble, France. The neutron
wavelength of 5\,\AA was selected and the collimator-sample and the
sample-detector distances were set to 5\,m. In all experiments the
sample together with the magnet is rotated and tilted, i.e. rocked
in order to move the magnetic diffraction peaks through the Ewald
sphere. The SANS patterns presented are the sum of intensities
through the whole rocking angle range. Similar measurements were
taken in the paramagnetic phase at 20\,K, and subtracted from the
data to obtain the magnetic scattering data.

\section*{Data Availability}
The measurement data
from ILL is publicly available under the ILL DOI-s
https://doi.ill.fr/10.5291/ILL-DATA.INTER-338 and
https://doi.ill.fr/10.5291/ILL-DATA.5-42-438. The rest of the data that support the findings of this study are available from the corresponding
author upon reasonable request.

\bibliography{mybib5}{}
\bibliographystyle{naturemag}

\textbf{Acknowledgements} We thank {\'A}. Butykai for stimulating
discussions. This work was supported by the DFG via the
Transregional Research Collaboration TRR 80 From Electronic
Correlations to Functionality (Augsburg/Munich/Stuttgart), by the
Hungarian National Research, Development and Innovation Office NKFIH, ANN 122879 BME-Nanonotechnology FIKP grant (BME FIKP-NAT), by
the Swiss National Science Foundation via Grants No. 200020-159893,
the Sinergia network Nanoskyrmionics (Grant No. CRSII5-171003), and
the NCCR Quantum Science and Technology (QSIT).

\textbf{Author Contributions} V.T. synthetized the crystals; B.G.,
A.M., S.P., M.P. performed and analysed the torque measurements;
K.G., P.L. performed and analysed the pyro- and magneto-current
measurements; B.G.Sz., S.B., J.S.W., R.C., I.K. performed and
analysed the SANS measurements; S.G., S.W. performed and analysed
the magnetization measurements; I.K. wrote the manuscript; I.K.,
M.P., S.B. planned the project.

\textbf{Competing interests} The authors declare no competing
financial interests.

\textbf{Additional information} Supplementary Information is available for this paper. Correspondence and requests for materials should be addressed to K. G.. Reprints and permissions information is available at www.nature.com/reprints.


\end{document}